\documentclass[a4paper,fleqn,usenatbib]{mnras}

\usepackage{newtxtext,newtxmath}

\usepackage[T1]{fontenc}
\usepackage{ae,aecompl}

\usepackage{graphicx}	
\usepackage{amsmath}	
\usepackage{amssymb}	
\newcommand{\quotes}[1]{,,#1''}

\title[Water transport throughout the TRAPPIST-1 System]{Water transport throughout the TRAPPIST-1 system: the role of planetesimals}

\author[V.  \DJ o\v sovi\' c et al.]{
Vladimir \DJ o\v sovi\' c,$^{1}$\thanks{E-mail: vladimir\_djosovic@matf.bg.ac.rs}
Bojan Novakovi\' c,$^{1}$
Branislav Vukoti\' c,$^{2}$
Milan M. \' Cirkovi\' c$^{2}$
\\
$^{1}$Department of Astronomy, Faculty of Mathematics, University of Belgrade, Studentski Trg 16, 11000 Belgrade, Serbia\\
$^{2}$Astronomical Observatory of Belgrade, Volgina 7, 11000 Belgrade, Serbia
}

\date{Accepted XXX. Received YYY; in original form ZZZ}

\pubyear{2019}

\begin{document}
\label{firstpage}
\pagerange{\pageref{firstpage}--\pageref{lastpage}}
\maketitle

\begin{abstract}
Observational data suggest that a belt of planetesimals is expected close to the snow line in protoplanetary disks. Assuming there is such a belt in TRAPPIST-1 system, we examine possibilities of water delivery to the planets via planetesimals from the belt. The study is accomplished by numerical simulations of dynamical evolution of a hypothetical planetesimal belt. Our results show that the inner part of the belt is dynamically unstable and planetesimals located in this
region are quickly scattered away, with many of them entering the region around the planets. The main
dynamical mechanism responsible for the instability are close encounters with the outermost planet
Trappist-1h. A low-order mean-motion resonance 2:3 with Trappist-1h, located in the same region, also contributes to the objects transport.
In our nominal model, the planets have received non-negligible amount of water, with the smallest amount of $15$\% of the current Earth's water amount (EWA) being delivered to the planet 1b, while the planets Trappist-1e and Trappist-1g have received more than $60$\% of the EWA.  
We have found that while the estimated efficiency of water transport to the planets is robust, the amount of water delivered to each planet may vary significantly depending on the initial masses and orbits of the planets.
The estimated dynamical \quotes{half-lives} have shown that the impactors' source region should be emptied in less then 1~Myr. Therefore, the obtained results suggest that transport of planetesimals through the system preferably occurs during an early phase of the planetary system evolution.
\end{abstract}



\section{Introduction}

The TRAPPIST-1 system was discovered in 2016, using TRAnsiting Planets and
PlanetesImals Small Telescope \citep{Gilon1}. The system is around 12~pc away from
the Earth, and consists of seven Earth-like planets \citep{Gilon2}. Even though
TRAPPIST-1 is a very old system, aged 7.6 $\pm$ 2.2~Gyr \citep{Burazer}, it is a
very compact system, with all the planets orbiting within 0.07~au. Stability of this
compact system is explained by a resonant chain that possibly includes all seven
planets \citep{Tamayo2017,Luger}. 

Habitable Zone (HZ), when considered in the context of planetary science, is an area around host star where temperature allows liquid
water to exist on planetary surface \citep{gonzales}. From anthropocentric
viewpoint, the existence of liquid water at planetary surface at fairly wide range of
atmospheric compositions and pressures is required for growth and development of
life. For this reason, it is important to determine whether or not there is a
liquid water on at least some of the planets in TRAPPIST-1 system, or at least
whether there could have been significant amounts of it in the past. The case of Mars, which is often regarded as fully habitable during the early Noachian age and perhaps marginally habitable today, shows that even water locked in subsurface deposits or permafrost counts toward the habitability index \citep{vestal}.

Water could be incorporated in planets during their formation or be delivered at
later stages of planetary system evolution. Since the planets located inside
habitable zones are likely formed inside a snow line, and thus are not made of
water-rich materials \citep{HZ1, HZ2}, water at their surfaces could have been
transported from outer parts of a planetary system. For a long time only postulated, the existence of such water reservoirs in the outer regions of planetary systems in the course of their formation has been also observationally confirmed \citep[e.g.][]{Hogar,Kama}. In this case,
water could be reimbursed at subsequent stages of evolution of the planetary systems,
during their migration stages \citep{tsiganis2005,migracija}. On the other hand, some papers
\citep{Colemanetal2017, Colemanetal2019, Scho19} suggested that planets would form
outside of snow line, and then migrate inward, thus accumulating higher water
fractions. However, in the latter scenario, the water on some TRAPPIST-1 planets would likely
be lost due to an XUV emission from Ultra Cold Dwarf star \citep{waterloss} through whole lifetime of planetary system. Still, water on them could be delivered after disk phase, at even later
stages of evolution of the planetary system, during impacts of planetesimals
originating outside of the snow line, thus containing volatile materials
\citep{Kral,Cuntz2018}.

For the planets in TRAPPIST-1 system, especially for those located inside the
habitable zone, it is
important to estimate the origin and amount of their water content.
\citet{Papa} found that three out of the seven planets  in this system,
namely Trappist-1e, Trappist-1f, and Trappist-1g, are in the host star
habitable zone. Additional analyses of the tidal heating \citep{tide}, and climate
modeling for planets in TRAPPIST-1 \citep{climate}, suggest that the most habitable
planet in this system should be Trappist-1e.  

The first indication of the existence of water at the surfaces of planets
Trappist-1b and Trappist-1c, comes from the Hubble Space Telescope
observations \citep{devit}. Moreover, using numerical simulation of transit
timing variations, \citet{Grimm} successfully determine densities of the planets in
TRAPPIST-1 system, and therefore planetary masses as well. Furthermore, based on
these results, and known diameters of the planets obtained by transit method
\citep{Gilon1, Gilon2}, \citet{Grimm} also estimated composition of each of the
planets. They suggested that the planets Trappist-1c and Trappist-1e are almost
completely rocky, while on Trappist-1b, 1d, 1f, and 1g, the existence of a surface
envelope is possible, either in a form of water-ice, ocean or an atmosphere.

It is well known that objects with about a kilometer in diameter are residuals of
planet formation \citep{formiranje}, regardless of the mechanism of their creation.
The existence of planetesimal belt similar to the one in Solar System is confirmed
observationally in some other planetary systems as well \citep{ppdisk}. Such objects
could have high influence on their neighborhood in many different ways, including
transport of materials from one part of the system to another. A well-known example
is the transport of water in our Solar System \citep{Obrajan}, but a similar
approach has been applied also to other planetary systems
\citep{2020tnss.book..331D, Frantseva2020}.

Many small objects in the inner Solar System are known to contain some amount of
water. The largest object in asteroid belt, (1)~Ceres, likely contains at least
$20\%\,$-$\,30\%$ water by mass \citep{McCord,
Castillo-Rogez_McCord2010,DownMission}. The water-ice has been detected at the
surface of asteroid (24)~Themis  \citep{Campis, Rikvin}, while the so-called
main-belt comets (MBCs), a subgroup of active asteroids, are expected to contain
significant amount of water \citep{komet,Snodgrass2017}. The association of MBCs to
primitive collisional asteroid families implies that water-ice is almost everywhere
in the asteroid belt \citep{hsieh2018}. Finally, recent result from the NASA's
OSIRIS-REx mission suggests that hydrated minerals are widespread on near-Earth
asteroid (101955)~Bennu \citep{Hamilton2019}. Although this does not imply that
Bennu contains any water right now, it shows that its parent body, probably a larger
asteroid than Bennu, did contain water. These facts show that water could
be widespread even among the objects in asteroid-like extrasolar planetesimal
belts.

A possibility to replenish planetary water reservoirs in TRAPPIST-1 system via
impacts by water-rich
planetesimals was a subject of two recent studies.
A water transport from the outer part of TRAPPIST-1 system was studied by
\citet{Kral}. These authors assumed existence of planetesimals on cometary-like
eccentric orbits in the system, and analyzed a potential effect of these objects for
water delivery and formation of the secondary atmospheres. They showed that
significant amount of water could be transported in such a way and that the
secondary atmospheres could be formed. The eccentric orbits however are not a
natural outcome of planetesimal formation process \citep{yang}, but may be produced
during a subsequent dynamical evolution. Still, such dynamical evolution does
not necessarily occurred in the TRAPPIST-1 planetary system. 

A similar study was performed also by \citet{Dencs2019}. These authors studied
the amount of water delivered to the planets from a water-rich asteroid belt located
just beyond the snow line, and also found
that this is a plausible way to deliver the water to the planets. However,
\citet{Dencs2019} assumed a presence of an additional planet in the TRAPPIST-1
system, embedded in the asteroid belt. Although the existence of such a planet
in the system cannot be ruled out, it is not discovered so far.

In more broader context, the dynamics of stellar encounters
\citep{bane_preporuka1, bane_preporuka2} can affect the orbital motion of
platenesimals, in particular those coming from the outer parts of a planetary
system. Both, systematic tidal effects of the Galactic disk, and stochastic
effects of encounters with spiral arms, giant molecular clouds, supernova remnants,
etc. represent major influences on habitability of planets within individual
planetary systems. This highlights the importance of galactic environment for
the habitability studies of the individual planetary systems. In turn, this means
that the results of the above described, and similar works (including the one presented in this paper), 
are also relevant for considering the Milky Way habitability timescales, such as in \citet{Dosovic}. 

In this work we have searched for an alternative scenario that would explain water
delivery to the planets in TRAPPIST-1 system. We have studied a long term dynamical
evolution of an asteroid-like planetesimal belt
within the currently proposed architectures of the system, and considered a potential amount of
water delivered from the belt.

\begin{figure*}
\begin{center}
\includegraphics[width = \linewidth]{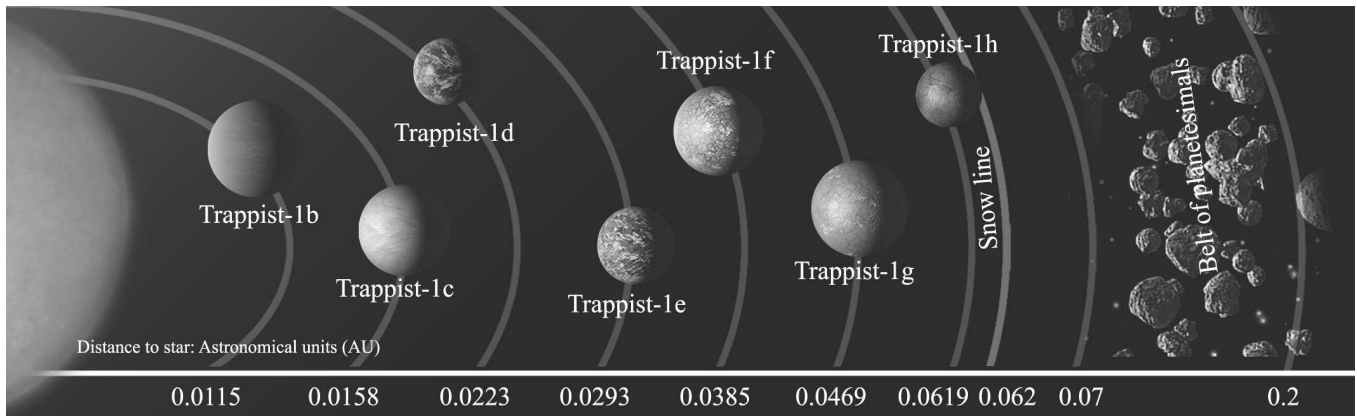}
\end{center}
\caption{Sketch of the TRAPPIST-1 system with seven planets. In addition to the
planets, our hypothetical planetesimal belt and the location of the snow line
(light blue) are shown. Credit: adapted from NASA/JPL-Caltech.}
\label{prsten}
\end{figure*}

\section{Numerical simulations}

\subsection{Numerical integrations}  \label{integratori}

In order to analyze a long-term dynamical evolution of planetesimals in the
TRAPPIST-1 system, we used numerical simulations. 

Observational data suggest that warm planetesimal belts often form in neighborhood
of a snow line \citep{Martin_evolucija_snowline}. 
As our goal is to simulate evolution of the hypothetical asteroid-like 
planetesimal belt, we assume that its inner edge approximately corresponds to the
location of
the snow line in the TRAPPIST-1 system. The radial extension of this belt is assumed
to correspond
to the extension of the asteroid belt in Solar system, but scaled to the size of the
TRAPPIST-1 system.

Location of the snow line in the TRAPPIST-1 system is determined using
methodology
described in \citet{ogihara} for optically thin disk. Based on their paper, we
found
that the snow line should be located at $0.062$~au, just beyond the
outermost planet Trappist-1h (see Fig. \ref{prsten}). Moreover, 
scaling from the size of the Solar System asteroid belt, we estimated that the
planetesimal belt should extend till $0.2$~au. We would like to note here that
without an extra planet outside $0.2$~au in the TRAPPIST-1 system, the disk of planetesimals could in principle spread beyond this limit. However, as results of our simulations presented in Section~\ref{sec:results} show, the planetary perturbations at this location should be negligible, and therefore, this unused part would not contribute anyhow to the amount of water delivered to the planets.

Orbital elements of test particles are distributed according to expected structure
of a somewhat excited protoplanetary disc \citep{knjiga}. The particles are
uniformly distributed in terms of orbital semi-major axis, from $0.07$ to $0.2$~au,
and in terms of orbital eccentricity, from $0.0$ to $0.2$. A half-normal
distribution is adopted for the orbital inclinations, above the mid-plane of
the protoplanetary disc, with a standard deviation of $1^\circ$. The remaining three
angular orbital elements are distributed randomly from $0$ to $360^\circ$. All
distributions are produced using generators from {\sc numpy} package.

The simulations are performed using a public domain {\sc Mercury}\footnote{The
package is written in {\sc FORTRAN~77}, and is available at
https://github.com/4xxi/mercury.} software package \citep{merkuri2}. This package
has been widely used to simulate dynamics of planetary systems \citep[e.g.][]{mer1,
mer2, mer3}. Indeed, \citet{mkritika} recently pointed out that some parts of
\emph{Mercury} code may not be properly described in \citet{merkuri2}, but
nevertheless, \citet{odgkritika} showed that this has no influence on accuracy and
performance of the integrators.

For the purpose of this study, we used two different integrators from the {\sc
Mercury} package.
The most of the simulations are performed using \emph{mixed-variable symplectic}
({\sc MVS}) integrator \citep{mvs}, but in order to have more reliable results, a
subset of orbits of planetary-approaching
planetesimals are numerically propagated using also \emph{hybrid} integrator
\citep{merkuri2}.

Numerical integrations of the TRAPPIST-1 system are performed using the dynamical
models that include seven planets as massive objects, and 20000 planetesimals as massless test particles.
In order to investigate how sensitive is the transport of water to the planets 
on the dynamical model, we use two different models of the TRAPPIST-1 system available in the literature \citep{pocetni2,Grimm}. From these models we adopted the initial conditions for planets, including their
orbital elements as well as physical characteristics such as masses and radii (see Table~\ref{tab:planete}).

\begin{table*}
 \caption{The initial orbits and masses for planets in TRAPPIST-1 system,
from two different models. In the table, letter G refers to model of \citet{Grimm}, and
letter W refers to model of \citet{pocetni2}}
   \label{tab:planete}
   \centering
\resizebox{\textwidth}{!}{

\begin{tabular}{lllllllllllllll}
Planet & \multicolumn{2}{c}{b}                         & \multicolumn{2}{c}{c}                         & \multicolumn{2}{c}{d}                         & \multicolumn{2}{c}{e}                         & \multicolumn{2}{c}{f}                         & \multicolumn{2}{c}{g}                         & \multicolumn{2}{c}{h}                         \\
Model   & \multicolumn{1}{c}{W} & \multicolumn{1}{c}{G} & \multicolumn{1}{c}{W} & \multicolumn{1}{c}{G} & \multicolumn{1}{c}{W} & \multicolumn{1}{c}{G} & \multicolumn{1}{c}{W} & \multicolumn{1}{c}{G} & \multicolumn{1}{c}{W} & \multicolumn{1}{c}{G} & \multicolumn{1}{c}{W} & \multicolumn{1}{c}{G} & \multicolumn{1}{c}{W}& \multicolumn{1}{c}{G} \\
    \hline
    Mass [$M_\oplus$] & 0.790 & 1.017 & 1.630 & 1.156 & 0.330 & 0.297 & 0.240 &
0.772 & 0.360 & 0.934 & 0.566 & 1.148 &
0.086 & 0.331\\[2pt] 
 Planetary radius [$R_\oplus$] & 1.086 & 1.121 & 1.056 & 1.095 & 0.772 & 0.784 & 0.918 &
0.910 & 1.045 & 1.046 & 1.127 & 1.148 & 0.715 & 0.773\\[2pt]
    Semi-major axis [au] & 0.01111 & 0.01155 & 0.01522 & 0.01582 & 0.02145  &
0.02228 & 0.02818 & 0.02928 & 0.03710 & 0.03853
& 0.04510 & 0.04688 & 0.05960 & 0.06193\\[2pt]
    Eccentricity & 0.019 & 0.006 & 0.014 & 0.007 & 0.003 & 0.008 & 0.007 & 0.005 &
0.011 & 0.010 & 0.003 & 0.002 &
0.086 & 0.006\\[2pt]
    Inclination [$^\circ$] & 89.393 & 90.000 & 89.626 & 90.000  & 89.866 & 90.000  &
89.754 & 90.000 & 89.694 & 90.000  &
89.707 & 90.000  & 89.814 & 90.000 \\[2pt]
    Longitude of ascending node [$^\circ$] & 307.0 & 0.0  & 201.5 & 0.0  & 338.7 &
0.0  & 322.2 & 0 .0
& 199.2 & 0.0  & 190.4 & 0.0  & 269.9 & 0.0 \\[2pt]
    Longitude of pericenter [$^\circ$] & 0.0 & 336.9 & 40.0 & 282.5 & 80.0 & -8.7 &
110.0 & 108.4 & 160.0 & 368.8 &
195.0 & 191.3 & 230.0 & 338.9\\[2pt]
    Mean anomaly [$^\circ$] & 344.0 & 203.1 & 232.9 & 69.9 & 371.0 & 173.9 & 185.5 &
347.9 & 4.5 & 113.6 & 185.9 & 265.1
& 93.2 & 269.7\\[2pt]
    \hline
   \end{tabular}}
  \end{table*}

The time span of numerical integrations is determined using a time scale in the
Solar System which is in most cases long enough to trace dynamical evolution of
small bodies. This time scale was found to be around 10~Myr \citep[see
e.g.][]{10Mgod}. As the relevant time scale corresponds to a number of orbital
periods, we found that a typical period of an asteroid in the Solar system
asteroid belt is about 20 times longer than a period of an object from our assumed
planetesimal belt. Therefore, in the numerical simulations of the TRAPPIST-1 system
we have propagated the orbits over a time span of $0.5$~Myr.

As already mentioned above, the orbits of 20000 planetesimals are initially
propagated using the {\sc MVS} integrator, and their orbital elements are sampled
every $50$~yr. These results are then analyzed for potential close encounters
between planetesimals and planets. The particles
that passed at a distance smaller than $0.1 R_{\mathrm{H}}$ (where
$R_{\mathrm{H}}$ is a Hill's radius of
a corresponding planet)\footnote{Due to numerous close encounters in the system, recording these encounters is highly memory demanding. In this respect, our choice to consider only encounters within $0.1 R_{\mathrm{H}}$ is found to provide a good compromise between accuracy and resource consumption.} from any of the planets, as well as those whose perihelion
distance at any time of the integration was smaller than the aphelion distance
of the most distant planet in the system, namely Trappist-1h, are identified and
their orbits are propagated again. These second integrations are performed using the hybrid
integrator with an adaptive time-step, and two sets of outputs are recorded: the
orbital elements of planetesimals, and the parameters of close approaches. Since
during a single close encounter, any planetesimal could, in principle, pass
several times inside the $0.1$ Hill's radius of the planet, only the data on
the closest approach is kept.

\subsection{Mass of the planetesimal belt}

The above adopted number of 20000 planetesimals is an arbitrary number selected to
be large enough for
statistical purposes, but also small enough to avoid highly time-consuming
simulations. It is however
very important to estimate a realistic number of objects expected in a belt similar
to the planetesimal 
belt assumed in this work.

In order to find a mass of the planetesimal belt, we need to find mass of the whole
protoplanetary disc. For this purpose, we assumed that mass of the disk $M_{\rm disk}$ represents $1\%$ of the mass of the host star $M_*$, i.e. $M_{\rm disk} = 0.01 M_*$, and dust to gas ratio in the disk of $f = 0.01$ \citep{knjiga}, is adopted. Based on this, we estimated a total mass of solids in the disk to be $M_{\rm solid} = 10^{-2} M_{\rm disk}$ = $10^{-4} M_*$. 


Using the mass of solids in the disk, adopting a surface density profile in the form
\begin{equation}
\Sigma = \Sigma_0 r^{-\beta}, 
\end{equation}
and assuming an exponent $\beta = 1.5$, as in the Solar System \citep{stepen}, the
mass of the planetesimal belt $M_\mathrm{belt}$ could be determined by integrating
the above equation from the inner to the outer edge of the belt. This yields the
following expression:
\begin{equation}
M_\mathrm{belt} = M_{\mathrm{solid}}  \frac{\frac{1}{r_1^{0.5}} -
\frac{1}{r_2^{0.5}}}{\frac{1}{a_1^{0.5}} - \frac{1}{a_2^{0.5}} },
\end{equation}
where $r_1$ and $r_2$ are inner and outer bounds of the planetesimal belt, respectively, while $a_1$ and $a_2$ represent borders of the whole Trappist-1 protoplanetary disc, adopted to be $0.01$ and $200$~au, respectively. 

The inner border of protoplanetary disc is taken
to be inside the orbit of the innermost planet Trappist-1b. The value of the outer
border is selected based on protoplanetary disc models and observations
\citep{ppdisk}, and is large enough to ensure that only negligible mass fraction of
the disk may be found outside this limit. With these assumptions, we estimated
the mass of the planetesimal belt to be $M_\mathrm{belt} =
0.46~M_{\oplus}$, assuming that whole mass of solids available in this part of the disk is incorporated into the planetesimals.

The study of  \citet{rejmond} could be interpreted as suggesting a somewhat higher value of $\beta$ in the TRAPPIST-1 system, as compared to the Solar System one, since they find steeper density profiles correlated with more terrestrial planets present in simulations. In the same time, however, this would make the uncertainty in position of the outer edge of the planetesimal belt less important, since outermost parts would negligibly contribute to the total mass. However, we decided to use the same value as in Solar System for parameter $\beta$. With some other value for $\beta$ we would get different estimation for mass of planetesimal belt, thus only the scaling factor from number of test particles to the real number of planetesimals would be different.

The next step is to determine a size-frequency distribution (SFD) of planetesimals. The
SFD of small objects in a planetary system are usually assumed to be in a form 
\begin{equation}
N(>D) = N_{\mathrm{t}}D^{-\alpha},
\label{eq:sfd_belt}
\end{equation}
where $N_{\mathrm{t}}$ is the total number of objects and $D$ is given in kilometers. Therefore, in order
to determine the SFD of objects in the planetesimal belt (under the assumption that
there is no preferential selection by size), we should know the parameter $\alpha$.

\citet{jorgos} found that the exponent $\alpha$ of primordial cumulative size
distribution of objects in the Solar System asteroid belt should be $\alpha =
1.43$, for objects bellow $100$~km in diameter, and $\alpha = 2.5$ for objects
larger than $100$~km in diameter. 

If we assume that all test particles are spherical, total mass in planetesimal belt
$M_\mathrm{belt}$ could be expressed as:
\begin{equation}
M_\mathrm{belt} = \frac{1}{6}\rho \pi \sum_{i = 1} ^{N_{\mathrm{t}}} D_i^3,
\label{eq:mass_belt}
\end{equation}
where $\rho$ represents density of planetesimals, $N_\mathrm{t}$ total number of
planetesimals in our model and $D_i$ diameter of $i$-th planetesimal. For the
density of planetesimals we assume $2$~g/cm$^3$ , similar to the density of
water-bearing asteroid (1)~Ceres \citep{park2016}.

Combining the above-given Equations~\ref{eq:sfd_belt} and \ref{eq:mass_belt}, and
assuming that all the mass is concentrated in objects larger than $17$~km in
diameter, the mass of the planetesimal belt could be expressed as a function of
total number of planetesimals in a form: 
\begin{equation}
M_\mathrm{belt} = \frac{1}{6}\rho \pi \sum_{N = 1} ^{N_{\mathrm{t}}}
\Big(\frac{N}{N_{\mathrm{t}}}\Big)^{-3/\alpha}.
\end{equation}
Denoting with $\lfloor A \rfloor$ the largest integer number smaller than a real
number $A$, and substituting $\lfloor N_{\mathrm{t}} \cdot 100^{-1.43} \rfloor$ with
$N_{\mathrm{B}}$, we could write:
\begin{equation}
M_\mathrm{belt} = \frac{1}{6}\rho \pi \left[\sum_{N = N_{\mathrm{B}} + 1} ^{N_{t}}
\left(\frac{N}{N_{\mathrm{t}}}\right)^{-\frac{3}{1.43}} + \sum_{N = 1} ^{
N_{\mathrm{B}}} \left(\frac{N}{N_{\mathrm{t}} \cdot
100^{-\frac{2.5}{1.43}}}\right)^{-\frac{3}{2.5}}  \right].
\end{equation}

Solving the last equation iteratively, we found that there should be more than $2.84$
million planetesimals in the belt around TRAPPIST-1 star. Numerical simulation of
dynamical evolution with this number of objects would be computationally very
demanding. Therefore, we decided to work with 20000 test particles, and then to
linearly scale all the obtained results so that they correspond to the estimated
number of objects.

\section{Results and discussion}
\label{sec:results}

The numerical simulations described above are performed in order to analyze
dynamical evolution of the planetesimal belt in TRAPPIST-1 system, focusing
primarily on the influence of the planetesimals on the planets. In this respect, we
studied source regions of potential impactors, determined the impact rate and
its
evolution over time, and estimated the total amount of water that could be
potentially delivered to each planet.

As mentioned in Section \ref{integratori}, we run the simulations for two different sets of
initial conditions taken from \citet{Grimm} and \citet{pocetni2}. One of
the key differences between these two sets are in masses of the planets, with the masses from \citet{Grimm} being mostly larger than those from \citet{pocetni2}. This is particularly the case for the outermost planet 1h (see Table~\ref{tab:planete}). The
results of this work mostly rely on the simulations with the initial conditions from
\citet{Grimm}. However, we have also presented  the
results of the simulations initiated with the values from  \citet{pocetni2}, in
order to get a better insight into the robustness of the results.

\subsection{Dynamical evolution of the planetesimal belt}
\label{ss:dyn_evo}

The first step was to extract from 20000 test particles only those that have close
approach with at least one of the planets. In the first set of
simulations\footnote{First set of simulations was performed using 200 batches with
100 test particles each, i.e. 20000 particles in total.}, performed using {\sc MVS}
integrator, we found 1726 (about $8.6\%$) and 1709 (about $8.5\%$) of such particles, using the model of \citet{Grimm} and \citet{pocetni2}, respectively.
Very similar numbers obtained with two different models might be a hint that
results are not highly model sensitive.

\begin{figure}
 \begin{center}
\includegraphics[width=\linewidth, height=\textheight, keepaspectratio]{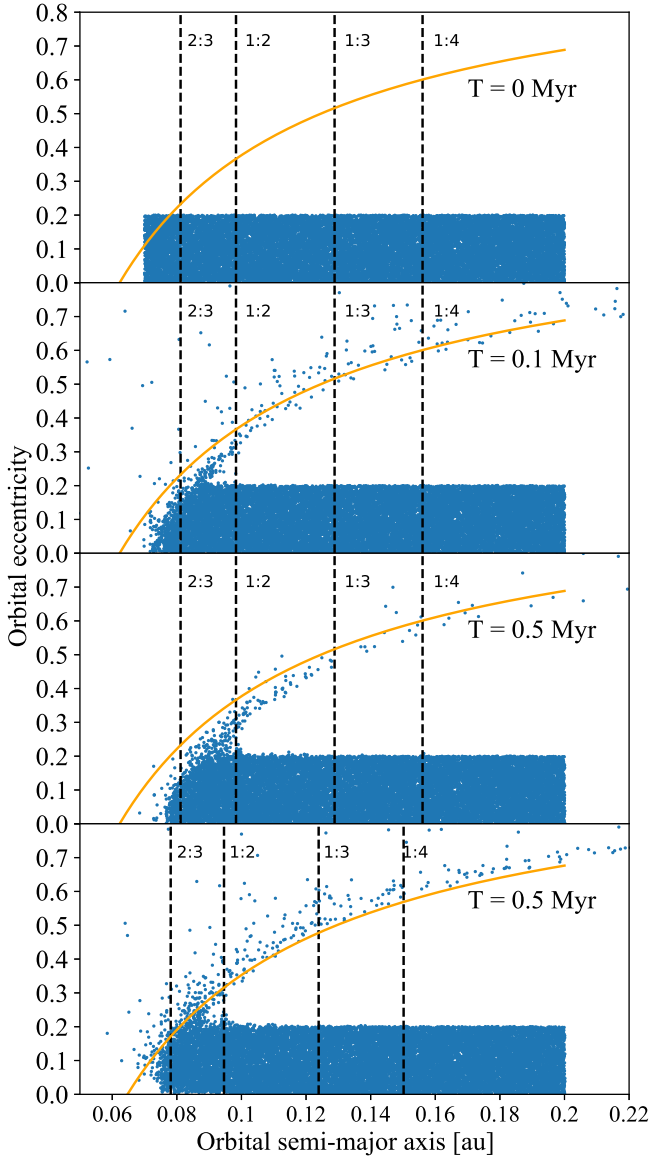}
 \end{center}
 \vspace{-10pt}
\caption{Snapshots from the dynamical evolution of the planetesimal belt. Figure shows evolution of 20000 test particles in the $a-e$ plane. The upper three panels show evolution within the model of \citet{Grimm}, at the beginning of 
the simulation (topmost panel), after $0.1$~Myr years (second panel), and after $0.5$~Myr 
(third panel from above). The lowermost panel presents the evolution after $0.5$~Myr of integration
obtained with the model of \citet{pocetni2}. Orange line marks location where the periastron distance of planetesimals is equal to the apoastron distance of the planet Trappist-1h. Black dashed line shows positions of the strongest mean-motion resonances with the most distant planet in the system, Trappist-1h.}
\vspace{-10pt}
 \label{Pocetak}
\end{figure}

Fig.~\ref{Pocetak} shows the snapshots from the simulations of the dynamical
evolution of the planetesimal belt within both models. The figure reveals that the most perturbed planetesimals during the dynamical evolution follow the line at which the planetesimals' periastron distance is equal to the apoastron distance of the planet Trappist-1h. The objects that reach this line become unstable due to repeated close encounters with planet 1h, and consequently leave the system quickly. We underline also that
the most strongly perturbed planetesimals are located close to the inner border of the belt,
while beyond about $0.1$~au perturbations practically vanish. General behaviors are very similar in both dynamical models. It should be however noted (see two bottom panels in Fig.~\ref{Pocetak}) that after 0.5~Myr of the evolution practically there are no planetesimals intersecting the orbit of the outermost planet 1h in the model of \citet{Grimm}, while there is still non-negligible fraction of such objects visible in
the simulation based on the model of \citet{pocetni2}. We believe this is a consequence of significantly larger mass of Trappist-1h in the model of Grimm and coauthors.

 Among the objects that have been removed before the simulation ends after $0.5$~Myr, $13$\% are ejected from the TRAPPIST-1 system, $8$\%  collide
with the central star, while remaining $79$\% impact the planets. If we exclude
planetesimals whose orbits initially intersect the orbit of the Trappist-1h, the numbers are somewhat different. In this case only $4$\% are ejected, $2$\% collide with the central star, and $94$\% of planetesimals impact the planets.

The further investigations are performed using the second set of the simulations, that
is produced using only 1726 (1709) objects that had close approaches with the planets in the first set of the simulations. Their orbits are propagated again for $0.5$~Myr, with the same initial conditions as for the first set, but using the hybrid integrator with an adaptive time step.

\begin{figure}
 \begin{center}
\includegraphics[width=\linewidth]{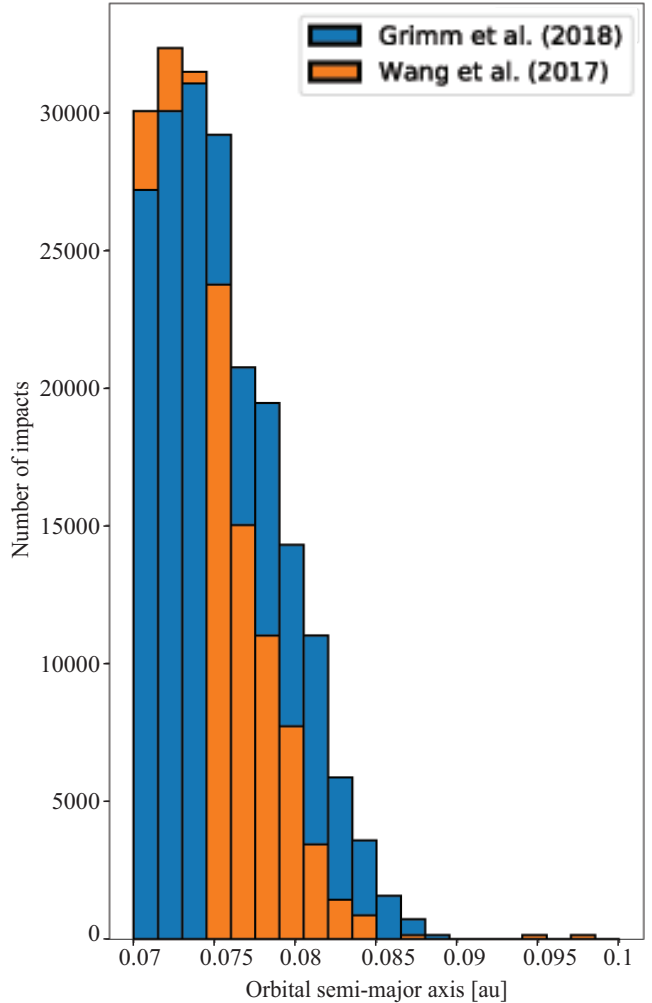}
 \end{center}
 \vspace{-10pt}
\caption{Number of objects impacting the planets in the TRAPPIST 1 system, as a
function of their initial semi-major axis, for the two different sets of initial
conditions (see legend). Note that the results are scaled to the expected real number of planetesimals in the TRAPPIST-1 system. Both distributions peak close to the inner
edge of the planetesimal belt.}
\vspace{-10pt}
 \label{hist_source_a}
\end{figure}

The results from the second set of integrations, presenting a number of impacts as a function of the distance from the host star, are shown in Fig.~\ref{hist_source_a}.
These results immediately point out the source region in terms of the initial
orbital semi-major axis, from which potential impactors are dominantly originating. It
could be easily seen in Fig.~\ref{hist_source_a} that almost all potential impactors are coming from the inner boundary of the planetesimal belt,
irrespective of the selected set of the planetary initial conditions. Some of the impactors were
initially placed on the orbits that intersect orbit of planet Trappist-1h, making
therefore their dynamical evolution controlled by direct gravitational interaction
with this planet. Close approaches with the planet 1h quickly increase orbital eccentricity of these particles, putting them on the orbits that intersect the orbits of the other planets as well.

\begin{figure}
\begin{center}
\includegraphics[width=\linewidth]{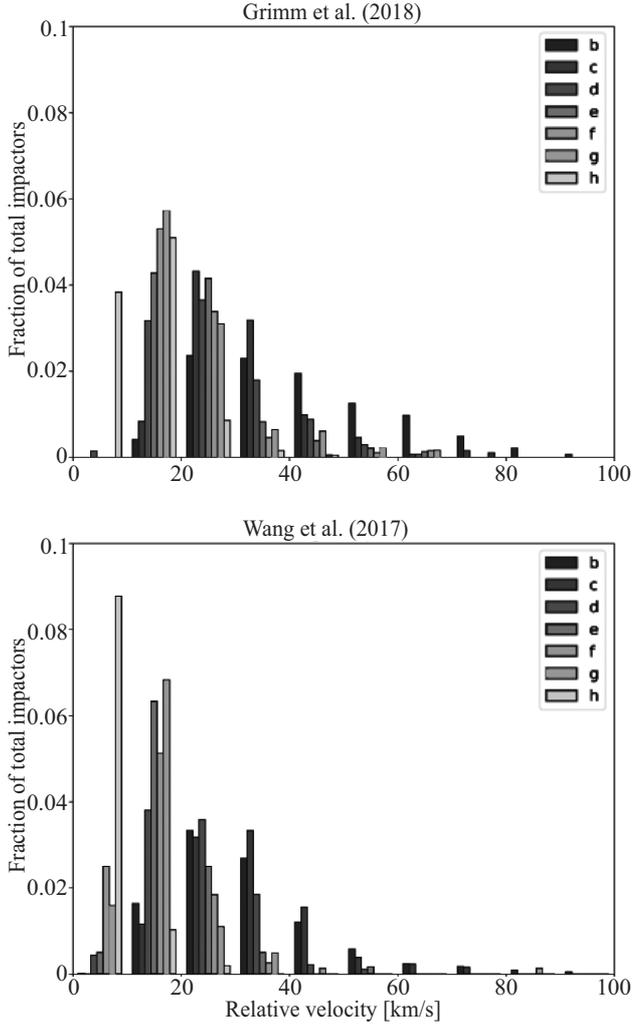}
 \end{center}
 \vspace{-10pt}
\caption{Histogram of relative velocities between planetesimals and planets in
TRAPPIST-1 system, at the moments of impacts. The upper panel shows results for
\citet{Grimm} model, while the lower panel presents results obtained using \citet{pocetni2} model. Each planet is presented with
different colour (see legend). Two panels correspond to the two labeled models.}
\vspace{-10pt}
 \label{hist_source_rel}
\end{figure}

Relative velocities at the time of close encounters may provide useful information
about impact conditions. Distribution of relative velocities with planets from the
TRAPPIST-1 system are shown in Fig.~\ref{hist_source_rel}. Since this system is very
compact, orbital velocities of the planets are very high, and for the closest planet
1b, this velocity is about $80$~km/s. When combined with not so small orbital
eccentricities of planetesimals that may impact the planets, this results in a
very high relative velocities. 

Distribution of relative velocities with planet 1b exhibits a single peak at about
$30$~km/s (Fig.~\ref{hist_source_rel}). It is useful to note here that \citet{Kral}
performed similar study, but assuming more distant planetesimal belt with very
eccentric trajectories, similar to the orbits of Long Period Comets in Solar System. In their
analysis of relative velocities, however, Kral and coauthors found two peaks
\citep[Figure~6 in ][]{Kral}. This fact is a consequence of different orbital
eccentricities of planetesimals assumed by \citet{Kral} and in this work.

\begin{figure}
 \begin{center}
\includegraphics[width=\linewidth]{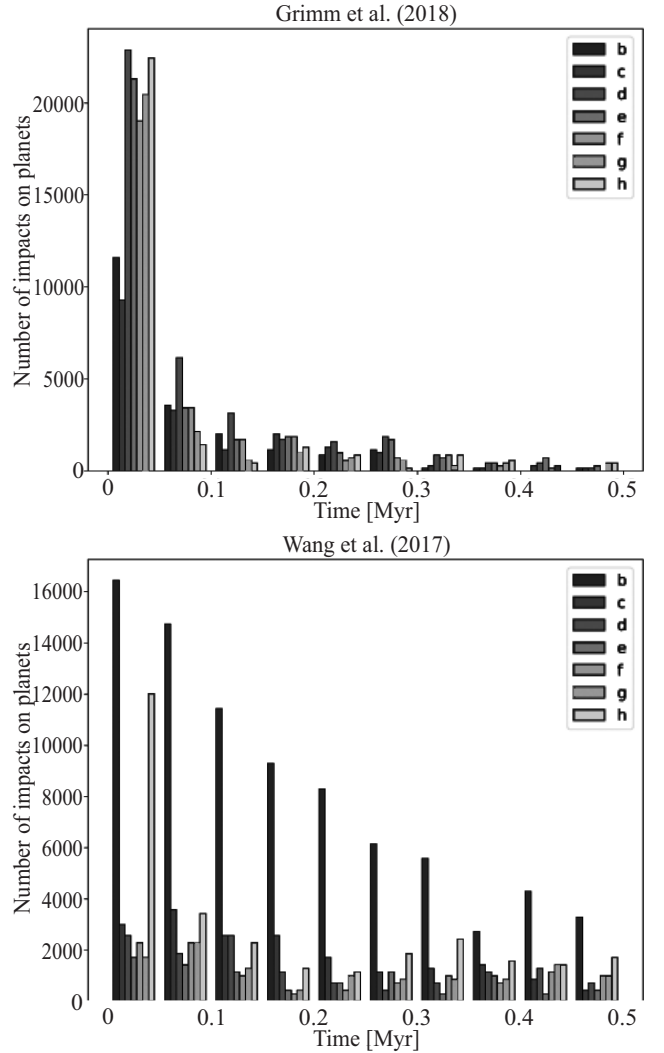}
 \end{center}
 \vspace{-10pt}
\caption{Time distribution of the number of planetesimal impacts to the planets
in the TRAPPIST-1 system. Two panels correspond to the two labeled models.}
\vspace{-10pt}
 \label{hist_vreme_d-h}
\end{figure}

A number of impacts on planets over the $0.5$~Myr long
time interval is given in Fig.~\ref{hist_vreme_d-h}. From this figure, a trend of
exponential decay of the number of close approaches over time can be noticed. This
indicates rapid depletion of the source regions of potential impactors. Fig.~\ref{hist_vreme_d-h} also shows a total number of impactors with individual
planets. The distributions of impacts on different planets obtained for two models are significantly different, as can be seen from
this figure. In the model of \citet{pocetni2}, planet Trappist-1b receives cumulatively by far the most of the impacts, while the rest of the planets are significantly less frequently impacted. On the other hand, in the model of \citet{Grimm} planets receive relatively similar number of impacts, with the two innermost planets being somewhat less frequently impacted.




%
%

In order to investigate in more details how flux of planetesimals towards each of
the planets fades over  time, we determine the so-called dynamical
\emph{half-life}\footnote{\emph{Half-life} is the time required for a given
population to decrease to half its initial value.} of impactor population.  To this
purpose the results shown in Fig.~\ref{hist_vreme_d-h} are fitted with a function of
the form
\begin{equation}
N(t) = N_0 \cdot e^{\frac{-\ln 2}{\tau}t}
\end{equation}
\citep[e.g.][]{half}, where $N$ is a number of impactors at an arbitrary time $t$, $N_0$ is an initial number
of impactors, and $\tau$ is a half-life.

The obtained results are given in Table~\ref{tab:half-life}. It is immediately clear
that dynamical lifetime
of impactors is  very short, with the half-lives for \citet{Grimm} model decreasing from about 207~kyr for the innermost planet 1b, to only about 146~kyr for the
outermost planet 1h. For \citet{pocetni2} model we found somewhat longer lifetimes of impactor populations, but still shorter than 1~Myr (see Table~\ref{tab:half-life}). Therefore, source regions of impactors will be almost
completely depleted on the timescale shorter than $1$~Myr. This suggests that impacts on
the planets in TRAPPIST-1 system from the planetesimal belt located near the snow
line are possible only for a short period of time in comparison to the age
of the system. These dynamical half-life times also imply planetesimals
transportation time scale of $\approx50$ kyr. This implies that the compact
planetary systems, such as TRAPPIST-1 can incorporate material from their outskirts
to the innermost planets in a relatively very short time interval.


\begin{table}
\caption{Dynamical half-lives of impactors for each planet in the TRAPPIST-1 system, and from two different dynamical models of the system.}
\label{tab:half-life}
\vskip4mm
\centering
\begin{tabular}{ccc}
\hline
Planet & \multicolumn{2}{c}{Dynamical half-life [kyr]} \\
  & \multicolumn{1}{c}{\citet{Grimm}} & \multicolumn{1}{c}{\citet{pocetni2}} \\
\hline
Trappist-1b                  & 207   & 378   \\
Trappist-1c                  & 191   & 425   \\
Trappist-1d                  & 183   & 447  \\
Trappist-1e                  & 166   & 462   \\
Trappist-1f                  & 161   & 458   \\
Trappist-1g                  & 151   & 853   \\
Trappist-1h                  & 146   & 143  \\
\hline
\end{tabular}
\label{tab:half-life}
\end{table}

The source region of potential impactors is crossed by one mean-motion resonance of
the first order, namely 2:3 with Trappist-1h. This resonance is located at
semi-major axis of $0.0812$~au, and may play an important role in the
transport of material from the belt to the planets. 
The mechanism at work here is similar as in the case of the 3:1 mean-motion
resonance with Jupiter in the Solar System \citep{Gladman1997Sci}. The 2:3
resonances with the planet 1h increases the orbital eccentricity of affected
objects till their periastron distance become small enough to allow close encounters
with the outermost planet 1h. Once this happens, the close encounters take the
leading role in transporting these objects towards the inner part of the system.

In order to maintain a flux of planetesimals towards the planets
in the TRAPPIST-1 system over longer timescales, an additional mechanism that would
inject new planetesimals in the
unstable regions is needed. In the Solar system, for km-sized objects, this role is
played by the non-gravitational Yarkovsky effect \citep{bottke2006}. This effect
changes the orbital semi-major axis of minor bodies, allowing many of them to reach
the resonances.

In order to estimate importance of the Yarkovsky effect in the TRAPPIST-1 system, we
used a model of \citet{vo2, vo1}, and estimated that for objects of 1~km in size an
expected semi-major axis drift rate
is about $da/dt = 1.5 \times 10^{-6}$ au/Myr. This value is about three orders
of magnitude smaller than
those obtained for the asteroid belt in the Solar system \citep{vok2015,nov2017},
which is due to a low luminosity of the Trappist-1 star compared to Solar
luminosity. Therefore, the Yarkovsky effect seems negligible in the TRAPPIST-1
system, and consequently an initial flux of planetesimals from the identified source regions
should decay very quickly, and should practically vanish in less than 1~Myr.
However, it should be noted that due to the compactness of the system, objects need to cross
comparatively smaller distance in order to reach unstable regions. Therefore, the Yarkovsky
effect might still be relevant for sub-kilometer bodies.

\subsection{Alternative source regions}

As discussed above, the mean-motion resonance 2:3 with Trappist-1h is located
inside the main source region of potential impactors, and may significantly
influence dynamical evolution of the planetesimal belt. Therefore, although a Jupiter-mass planet\footnote{Formation of a big Jupiter-mass planets 
around M dwarf stars is generally unlikely \citep{nema_Jup}} has not been discovered in the system \citep{Gilon2}, neither it is expected to exist \citep{Boss2017}, we found that first-order mean-motion resonance with a sub-Earth-mass planet are powerful enough to induce strong orbital
perturbations.

In order to understand better a potential role of other mean-motion resonances, here
we explore the effect of another first-order mean-motion resonance in the system, namely
the 1:2 resonance with Trappist-1h.

We note here however that this resonance is located at about $0.0983$~au, and that
we did not see
any obvious effect at this location in our numerical simulations (see
Fig.~\ref{Pocetak}). Nevertheless, we decided to investigate the role of the 1:2
resonance in detail. For this purpose, we adopted an approach developed by
\citet{Bojan2} to study diffusion among asteroid families.\footnote{Asteroid
families are
populations of asteroids that have very similar orbits, and are thought to be
fragments of past asteroid collisions \citep{Milani2014}.}

The diffusion is measured in space of two actions defined as \mbox{$J_1 \sim 1/2
\sqrt{a/a_\mathrm{p}}e^2$} and \mbox{$J_2 \sim 1/2 \sqrt{a/a_\mathrm{p}}\sin^2(i)$}~
\citep{Bojan2}, where $a$, $e$ and $i$ are semi-major axis, eccentricity and
inclination respectively, while index $p$ is used to denote orbital elements of
the planet, in this case 1h. 
Hence, we generate a new set of 100 test objects located inside the 1:2
resonance, and propagate numerically their orbits for 10 Myr, using the dynamical model from \citet{Grimm}. From these data we
have produced a time-series of $J_{i}$.
Finally, we assume a linear diffusion, and the so-called diffusion coefficients
$D(J_i)$ are obtained by linear fitting of time series of $\langle(\Delta
J_i)^2\rangle,(i = 1, 2)$, determined from the numerical integration for each action
$J_i$.
 
\begin{figure}
 \begin{center}
\includegraphics[width=\linewidth]{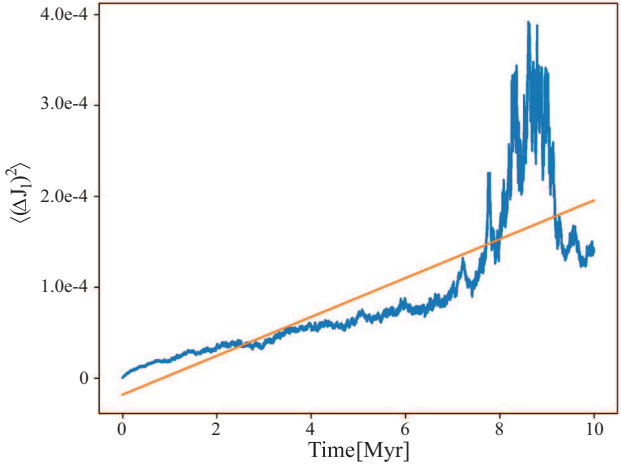}
 \end{center}
  \vspace{-10pt}
\caption{Time evolution of mean-square displacement in $J_1$ action for objects
inside the 1:2 mean motion resonance with Trappist-1h (blue curve), and its
corresponding linear fit (orange line).}
\vspace{-10pt}
 \label{J1}
\end{figure}

\begin{figure}
 \begin{center}
\includegraphics[width=\linewidth]{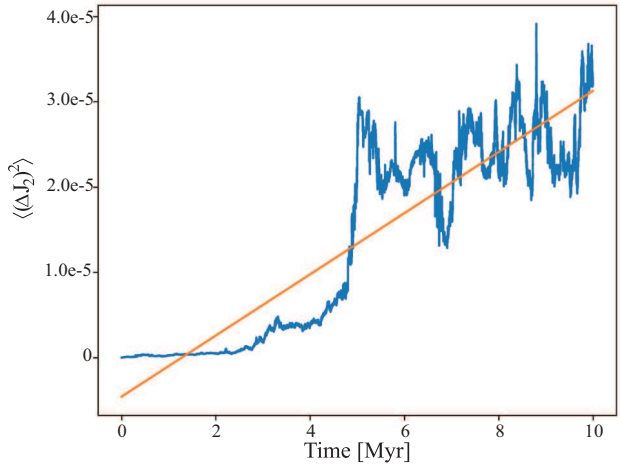}
 \end{center}
  \vspace{-10pt}
\caption{Same as on Fig.~\ref{J1}, but for action $J_2$.}
\vspace{-10pt}
 \label{J2}
\end{figure}

The obtained results reveal that the orbit diffusion process inside this resonance
is slow, but non-negligible. A significant change in $J_1$ occur on time scale $\sim 1$~Gyr
(Figs.~\ref{J1}). In other words, objects initially placed on eccentric orbits ($e \geq 0.2$) may evolve to the planet Trappist-1h crossing-orbit on the time scale of about $1$~Gyr. As shown in Fig.~\ref{J2}, diffusion is about an order of magnitude slower in $J_2$ action. Nevertheless, the diffusion observed in the $J_1$ action suggests that the 1:2 mean-motion
resonance might be a source of additional impactors after initial phase of
evolution of the belt, which takes 1~Gyr. Once the resonance remove all planetesimals trapped inside it, additional particles could be inserted by collisions near the borders of the
resonance \citep{collisions}, thus producing a new flux towards the planets, but of generally 
smaller objects.


\subsection{Water transport to the planets}

An important step in our analysis is to determine the amount of water transported to
the planets by planetesimals. To this purpose we used close encounters data
presented in Section~\ref{ss:dyn_evo}.
Specifically, if a distance between a planetesimal and a planet is found to be
smaller than the radius of the planet (see Table \ref{tab:planete}), we assume that an impact occurs. Next, we also assume that each planetesimal has a water content of $5$\% by volume, and
that during the impacts planetesimals transfer the whole amount of water to 
the planets, i.e. do not account for the potential water loss that could occur during
accretion \citep{Obrajan,Cieslaetal2015}.

The amount of water delivered to each planet is expressed as percentage of the
Earth's water amount (hereafter EWA). In this respect, let us recall that the amount of Earth's water in all aggregates is estimated to be 1,386,000,000~km$^3$ \citep{peter}.

\begin{figure}
 \begin{center}
\includegraphics[width=\linewidth]{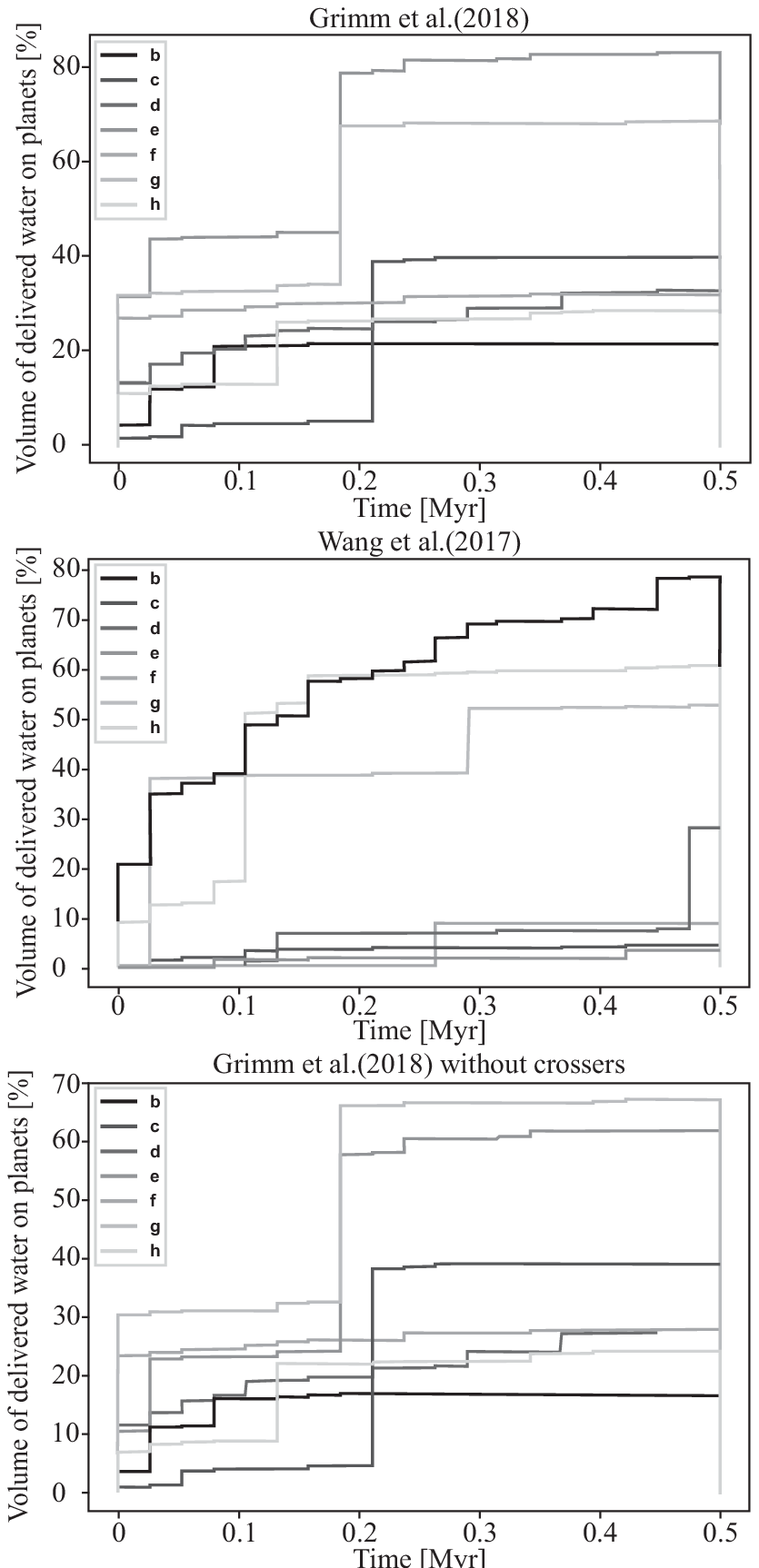}
 \end{center}
  \vspace{-10pt}
\caption{Cumulative amount of water delivered to planets as a function of time,
expressed as a fraction of the Earth's water volume. The results for  model of
\citet{Grimm} are shown in the upper panel, while model of \citet{pocetni2} is
shown in the middle panel. The lower panel shows the model of
\citet{Grimm}, with  the intersecting planetesimals excluded. Note that larger jumps in the amount of delivered water are associated to the impacts of big planetesimals. As the amount is estimated by scaling from the number of the simulated particles to the real number of objects in planetesimal belt, the timeline of water delivery may not be fully certain, but still the total amount should be realistic.}
\vspace{-10pt}
 \label{hist_volume}
\end{figure}

The cumulative amount of water delivered to the planets is shown in
Fig.~\ref{hist_volume}. The results obtained using the dynamical model by \citet{Grimm}
suggest that the largest amount of water could be delivered to the planets $1$e and $1$g.
Each of these two planets received more than 65\% of EWA. However, it is even more important
to note in these results that practically all the planets received non-negligible amount
of water, with the smallest amount of about 20\% of the EWA being delivered to the planet $1$b (see top panel in Fig.~\ref{hist_volume}). Excluding planetesimals that initially cross the orbit of Trappist-1h reduces slightly the amount of water delivered on the planets (bottom panel in Fig.~\ref{hist_volume}). Still, the main conclusions remain the same, and the
delivered amount of water is in excess of 15\% of the EWA for all the planets.


The results are different in the simulations performed with the initial conditions suggested by \citet{pocetni2}. In this case, the largest amount of water is delivered to the planets 1b, 1g and 1h,  in each case more than 50\% of Earth's water content, while the planets 1c and 1e received only a few percents of the EWA. Therefore, although generally transport of significant amount of water to the planets
in TRAPPIST-1 system via planetesimals is definitely possible, the results are model dependent to
some extent. It should be noted, however, that Trappist-1b is very close to the host star, and
therefore it is expected that any water delivered to this planet may exist only in
gaseous state. On the other hand, in the case of Trappist-1h, it is expected that
water delivered on this planet is frozen. The planet that receives more than 50\% of EWA, regardless of the applied set of initial conditions, is 1g. Interestingly, this is one of three planets located inside the habitable zone, \citep{Papa}, thus it is possible that at least some of its water may be in liquid state.

Recently several works performed composition and climate modeling of planets in the TRAPPIST-1 system.
\citet{Grimm} have analyzed the nature of planets in TRAPPIST-1 system, and
obtained so far the most precise densities of these planets. Regarding the composition, their results could be
summarized in the following. The planets 1c and 1e are likely rocky, while the
planets 1b, 1d, 1f, 1g, and 1h may have envelopes of volatiles in the form of thick atmospheres, oceans, or ice. Especially, these authors found that planets 1b, 1d, and 1g almost certainly are
volatile-rich. 
More recently, \citet{Unterborn2018RNAAS} performed more detail
modeling of planetary composition
in TRAPPIST-1 system and obtained results generally consistent with that of \citet{Grimm}. In
particular, \citet{Unterborn2018RNAAS} found that for planets 1b, 1d, and 1f, the results
strongly suggest the presence of planetary surface volatiles, but pointed out that the results are model dependent, and
more refined mass and radius measurements of the TRAPPIST-1 planets are needed for
better characterization.

Some climate models of TRAPPIST-1 planets suggest that Trappist-1e, if
it is today in synchronous rotation state and abundant in water, then this planet should always sustain surface liquid water at least in the sub-stellar region, whatever the atmosphere considered \citep{ocean2}. Also, \citet{climate} suggest that Trappist-1e may produce habitable surface temperatures beyond the maximum greenhouse distance.

It is interesting to compare our results with predictions based on the above-mentioned climate and composition models of planets in the TRAPPIST-1 system. In this respect, these models all predict that planet 1f should
contain a significant amount of water. The results of our simulations performed within the dynamical model from
\citet{Grimm} are in reasonably good agreement with the prediction by the climate and composition models, except maybe for planet 1b. However,
the results obtained using \citet{pocetni2} model predict only a small amount of water may be delivered to the planet 1f via planetesimals. For planet 1g, within the both dynamical models we found that this planet should be water-rich, in accordance with the results reported by \citet{Grimm} and \citet{Unterborn2018RNAAS}. The planet 1e is an interesting case. In our model with the initial conditions from \citet{Grimm} this planet receives a huge amount of water. This result is in disagreement with those by Grimm and coauthors, who found that 1e should be dry rocky planet. On the other hand, several other papers \citep{climate, ocean2} considered the possibility that the planet 1e may be tidally locked aqua-planet, in agreement with our results, regarding the water content. Still, using the model from \citet{pocetni2}, we found that planetesimals could deliver only a very modest amount of water to the planet 1e, but a large amount to 1b.
The results for 1e clearly illustrate that the current predictions are still not well constrained, and the situation is generally similar for all the planets.  Therefore, further theoretical and observational efforts are needed to improve the models and clarify these issues.

The discriminants between the predicted compositions require additional observational data, and new models that will involve different effects as well as their complex mutual interactions \citep[see e.g.][]{2020arXiv200501740B}. Nevertheless, our simulations provide useful constrains on possible water transport within the TRAPPIST-1 system. In general, we believe that the mechanism of water delivery to the planets in TRAPPIST-1 system investigated here is plausible, and it may explain the origin of water in at least some of these planets.

Also, we recall here that the water transport via planetesimal mostly occurs in an early phase of the planetary system, while the climate and composition models refer to present states of the planets. Therefore, these results refer to different points in time, and planets may undergo significant evolution (e.g. water loss) in the meantime. This might be an explanation for some of differences between our results and the climate and composition models predictions.
Of course, we cannot enter here the complex issue such is water retention, which may be
entirely different on planets around M dwarfs due to tidal locking, and strong XUV
flares \citep{habitat_dwarf}. In addition, the possibility of water retention should be examined against
the long term possibilities of water delivery, including the replenishment of the
planetesimal belt population with possible means
such as the stellar encounters and other possible effects related to the galactic
environment. This is of particular importance for the habitability studies of the
planetary systems that have  a relatively large estimated ages, such as the
TRAPPIST-1.

\subsection{Limitations of the water transport model}

The model of water transport presented here obviously has some limitations.

It is important to note that the $5\%$ water fraction in planetesimals assumed here 
is very conservative. 
It is based on what we see today in carbonaceous chondrites, which are the product
of the evolution of their parent bodies that formed from a mix of rock and ice a
billions of years ago. Therefore, it is likely that these parent bodies
contained significantly larger amount of water \citep{Cieslaetal2015}. On the other
hand, much longer history of the TRAPPIST-1 system allows for evolutionary processes
which have not been noticed or appreciated enough in the Solar system context.

Through analysis of the remains of meteorites, which are believed to be
asteroids, in the Solar system, it is concluded that they can have $5-20\%$ of
water \citep{procenat}. Therefore, the percentage of water in planetesimal
volume significantly depends on both, the distance between these objects and
their parent star in the moment of their birth, and on their later evolution. 

The most important caveat to this work seems to be the location of the snow line. A
fraction of volatile materials in planetesimals obviously depends on the location of
the snow line; in our model, shifting this line a bit further from the host star,
would imply that many planetesimals located close to the inner border of the belt
would have a significantly smaller water content. As argued in this work,
that the source region of
potential impactors in TRAPPIST-1 is confined in the narrow area, exactly at
the inner edge of the planetesimal belt, 
this may significantly affect the estimation of the amount of water delivered to the
planets. 

The material evaporation during planetesimal impacts should also be considered. It
is inevitable that this sort of impacts, and contacts with planetary
atmospheres, would cause a certain amount of material loss \citep{denim}. For
this study, the water transport during the impact is considered to be maximally
efficient, which is not the case in reality. The amount of evaporated material at
the moment of the impact depends on the impact velocity and the escape velocity from
the planet \citep{Obrajan}. Based on the impact simulation they conducted, mainly
for the Earth and Mars, \citet{denim} have concluded that about 20\% of material is
lost at the moment of impact. However, this is not the only issue - another issue is
whether some of the matter which is not lost in the sense of being ejected is
chemically dissociated. One could have water being vaporized, and while
formally not lost in these models, still being thermally or photo-dissociated,
with hydrogen being subsequently lost through outgassing at longer timescales (and
oxygen retained and bound in minerals).  

Dynamical evolution of planetesimal belt would be different if there is another
distant planet in TRAPPIST-1 system that is not discovered yet \citep{Dencs2019},
thus also we would expect more delivered water on planets if that is the case.

The mutual gravitational perturbations between the planetesimals are neglected in
our model.
However, planetesimals larger than about 500~km in diameter may significantly
influence the motion
of nearby objects \citep{DL2012,Nov2015}.  In this respect, we expect that taking
into account
these perturbations would somewhat increase the total flux of planetesimals towards
the planets,
resulting in more efficient water transport.

 Finally, the limit of the planetocentric distances of $0.1 R_{\mathrm{H}}$, that we considered here, is set somewhat arbitrary. However, as we are primarily interested in a number of planetary impacts, we considered the range around the planets from which many close approaches result in an impact. This could slightly affect our results, as some of the planetesimals passing at larger distances might impacts the planets as well. Therefore, for this reason our estimation of the total amount of water delivered to the planets could be slightly underestimated.

\section{Conclusions and Future work}

\subsection{Conclusions}

In this work, we have investigated dynamical evolution of the
hypothetical, asteroid-like, planetesimal belt 
in TRAPPIST-1 exoplanetary system. In particular, we addressed the role of the
planetesimals in the origin of water on planets in this system. 

The preceding analysis has led us to the following conclusions:
\begin{itemize}

\item There are dynamical mechanisms capable to transport significant amount of
material from the inner
edge of the planetesimal belt to the planets in TRAPPIST-1 system.

\item The obtained results suggest that the amount of water that could be delivered to the planets is significant. For individual planets it vary from $15$ to $80$\% of the Earth's water content.



\item The dynamical half-life of impactor population is short, at best a less than one million years. Therefore, the population of impactors decays quickly, and significant transport of
materials, including water transport, is possible only within the first several million years since the formation of the planetesimal belt.

\item The 1:2 mean motion resonance with planet Trappist-1h could be an additional source of impactors from the belt. It initiates a slow orbit diffusion that over a time-scale of $1$~Gyr may drive trajectories of affected planetesimals to planet-crossing orbits.

\item The non-gravitational effects in the TRAPPIST-1 system are negligible, and
therefore, there are no other efficient mechanisms to resupply new bodies inside the impactors' source regions.

\end{itemize}

\subsection{Future work}

The questions addressed here should be further investigated in order to test these
initial findings,
and to address some closely related problems. Let us mention a few of them.

Some parameters adopted in our model are not well-constrained, and it would be
therefore useful to test
wider parameter space. An example is a protoplanetary disc mass, and the gas to dust
ratio in disks.
In this work, we assumed that mass of the disk is $1$\% of the star mass
\citep{knjiga}, and the gas-to-dust ratio of 100. Still, different estimates could
be found in the literature. \citet{masamala} suggested that
mass of the disks may be somewhat smaller, in the range of $0.2-0.6$\% of the star
mass, while in their study of the TRAPPIST-1 system, \citet{Ormel} suggest the disk
mass of $4$\%. 

Our model did not account for possible removal of planetesimals in an early phase of
the system evolution,
as it may be caused, for instance, by planetary migration \citep{grand}, and that
likely occurred in the
TRAPPIST-1 system as well \citep{Colemanetal2019}. In this respect, it would be
interesting to analyze the amount of material transport assuming the mass of the
planetesimal belt proportional to the current mass of the Asteroid Belt in the Solar
System \citep{Levison2015}, that is significantly smaller than the one used in our
model presented here.

As we have argued above, the location of the snow line significantly affects the
estimation of the amount of water delivered to the planets. Therefore, it would be
very interesting in future work to test whether larger amount of water content in
icy planetesimals would be able to counterbalance potentially smaller impactor
source regions, due to possibly more distant location of the snow line in the
system.

The habitability studies of such compact planetary systems
of similar age should consider the possibilities of water retention together with the replenishment of the planetesimal belt by
mechanisms related to the interaction with the galactic environment. This should be investigated together with the possibility that 
close stellar flybys might destabilize the orbits of planets or alter 
the flux of small bodies towards the habitable zone of the planetary 
system.

Finally, from the astrobiological point of view, some recent
works suggest that, despite being located inside the habitable
zone \citep{KralHZ}, planets Trappist-1e, Trappist-1f and Trappist-1g
may not be good habitats
because of the extreme ultraviolet radiation from M dwarf host star \citep{EUV}, as
well as destructive atmospheric effects of super-flares \citep{howard}. Only forthcoming searches for biosignatures, 
however, coupled with
better constraints on the age of TRAPPIST-1, will shed some light on
the astrobiological status of this fascinating system.

\section*{Acknowledgements}
The authors would like to thank the referee for the valuable comments which helped to improve the manuscript. We acknowledge financial support from the Ministry of Education,
Science and Technological Development of the Republic of Serbia: V\DJ~and BN through
the project ON176011 "Dynamics and kinematics of
celestial bodies and systems", and BV and MM\'C through the project
ON176021 "Visible and invisible matter in nearby galaxies: theory
and observations".

\section*{Data availability}

The data underlying this article will be shared on reasonable request to
the corresponding author.


\label{lastpage}

\end{document}